\documentclass{revtex4}  
 
\usepackage{graphicx} 
\begin{document} 
\bibliographystyle{revtex} 
 
 
\title{Cosmological Parameters, Dark Energy and Large Scale
Structure\footnote{Invited review for the Symposium
``The Future of Particle Physics'', Snowmass 2001.}} 
 
 
 
\author{Susana E. Deustua} 
\affiliation{Lawrence Berkeley National Laboratory,
Berkeley, CA 94720; sedeustua@lbl.gov} 
 
\author{Robert Caldwell} 
\affiliation{Dept. of Physics \& Astronomy, Dartmouth College,
Hanover NH 03755; Robert.R.Caldwell@dartmouth.edu} 

\author{Peter Garnavich} 
\affiliation{Dept. of Physics, University of Notre Dame,
Nieuwland Science Hall 213, IN 46566-5670; pgarnavi@nd.edu} 
 
\author{Lam Hui} 
\affiliation{Dept. of Physics, Columbia University,
538 W. 120th St., New York, NY 10027; lhui@astro.columbia.edu} 
 
\author{Alexandre Refregier} 
\affiliation{Institute of Astronomy, Madingley Road,
Cambridge, CB3 OHA, UK; ar@ast.cam.ac.uk}

 
 
\begin{abstract} 

We review the current status of cosmological parameters, dark
energy and large-scale structure, from a theoretical and observational perspective. We first present the basic cosmological parameters and discuss how they are measured with different observational techniques. We then describe the recent evidence for dark energy from Type Ia supernovae. Dynamical models of the dark energy, quintessence, are then described, as well as how they relate to theories of gravity and particle physics. The basic theory of structure formation via gravitational instability is then reviewed. Finally, we describe new observational probes of the large-structure of the universe, and how they constrain cosmological parameters.

\end{abstract} 
 
\maketitle 
 


\section{Introduction}	 
 
In what is now a classic story, lies the foundation of 21st century cosmology.  In 1917 Einstein formulated the field equations which suggested a dynamic universe, in contradiction to the available data which supported a static universe.  Einstein thus added his now famous ``fudge factor'', the cosmological constant.  Following Edwin Hubble's 1926 paper in which he suggested an expanding universe, Einstein removed the cosmological constant.  The subsequent history of cosmology has leapfrogged between experiments to measure the cosmological parameters (expansion rate, mass density, deceleration rate, geometry and now energy density) and theories to organize them.   
 
However, the last few years have been very exciting for cosmologists.
Cosmic microwave background experiments, observations of Type Ia
supernovae, large scale surveys of galaxies and clusters and the
Hubble key project have yielded unhitherto precise measurements of the
key cosmological parameters.  Furthermore, these new values have
provided a concordance model, a flat universe with a significant
component of ``dark energy'' which accelerates the expansion in the
current epoch, which challenges our fundamental theories of particle
physics.
 
In this report, we  summarize the current status of the cosmological parameters (the Hubble constant, mass and vacuum energy density (dark energy)), quintessence and large scale structure; and their import on modern cosmology. 
 
\section{Cosmological Parameters: The Hubble Constant} 
 
The Hubble constant is a fundamental parameter 
of modern cosmology but its value has remained 
very uncertain despite many experiments which 
claimed small errors. Recently, a major effort to use 
the Hubble Space Telescope (HST) to measure Cepheid variable 
stars in galaxies well beyond the local group was 
completed by the Distance Scale Key Project and 
has yielded a value of $72\pm 7$~km$\;$s$^{-1}$ \cite{freedman2001}. 
This comprehensive study used a number of tertiary distance 
indicators, such as Tully-Fisher and Type~Ia supernovae (SNIa) 
to reach the Hubble flow with consistent results. The value 
of 72~km$\;$s$^{-1}$ is also in accordance with an independent 
method employing surface brightness fluctuations of galaxies \cite{ajhar}.
Still, there remains another group using Cepheids 
with HST which continues to obtain a significantly lower 
value of $59\pm 6$~km$\;$s$^{-1}$ \cite{parodi}. 
 
The top of the distance ladder appears to be more secure 
than the middle or bottom rungs. From studies in 
the Hubble flow ($0.01<z<0.1$), the peak brightness 
of SNIa appear to be a function of a single parameter 
characterized by their light curve shape \cite{phillips}. SNIa can 
provide a distance good to 7\%\ to any galaxy that 
has a well-observed SNIa. And this technique can 
be extended to 'peculiar' SNIa, thus removing the need 
for selection criteria that might induce systematic error \cite{garnavich}. 
The major sources of uncertainty in the Hubble constant 
are currently the unknown effect of metallicity variations 
on the Cepheid period-luminosity relation and the 
continuing debate on the distance to the Large Magellanic 
Cloud (\cite{jha}). 
 
Other methods to determine the Hubble constant such 
as applying the Sunyaev-Zeldovich effect to galaxy 
clusters (\cite{mason}) or using the time delays in gravitational 
lens images (\cite{williams}) show promise but are currently limited 
by large systematic uncertainties. 
 
\section {Cosmological Parameters: Dark Energy} 
 
\subsection {Type Ia Supernovae and the discovery of the Dark Energy} 
 
Type Ia supernovae (SNe Ia) are very bright,  calibratable  standard candles.  Upon explosion they release some 10$^{51}$ ergs, outshining a typical galaxy. This makes them ideal tools for cosmology since they can be detected to high redshifts.  The actual measurements are fundamentally straightforward:  brightness and redshift.  The former is obtained with imaging data, and the latter from spectroscopy.   The relationship between the observed apparent brightness and the redshift of standard candles depends on the cosmological parameters mass density, vacuum energy density and curvature. 
 
Two groups, the Supernova Cosmology Project (SCP) and the High Z Team (HZT) have programs to observe high redshift SNe at $z > 0.3$.  To date the SCP and the HZT have together discovered over 100  SNe Ia.   In 1998 both teams independently announced their results of an accelerating universe, suggesting the presence of a vacuum energy or dark energy component to the total mass-energy budget(\cite{perlmutteretal1999}, \cite{riessetal1998}.
    
The standard method pioneered by the SCP for discovering high redshift SNe 'on demand', requires two epochs of observations spaced about 3 weeks apart, just after and just before consecutive new moons on a 4m or larger class telescope.   This cadence allows the discovery of a batch of SNe before they reach maximum brightness, and, importantly, enables spectroscopic and photometric follow up with scheduled observations.  Supernova type is confirmed and redshift determined for each candidate SNe with spectroscopy at peak brightness obtained with large aperture telescopes, e.g.  Keck I and II, the VLT.  Follow up photometry at regular intervals are obtained for a period of roughly 60 days around peak to determine the light curve shape in order to normalize the peak magnitude by correcting for the brightness-light curve width relation (see Fig. \ref{lightcurve}).

\begin{figure}
\vskip 0.25 in
\caption{Top panel: lightcurves for low redshift type Ia supernovae from the Calan-Tololo Survey showingthe diversity in brightness and light curve width.  Bottom panel:  the same SN lightcurves after correcting
for the magnitude-width (empirical) relation by ''stretching'' the time axis.  They now lie on the same curve.}
\label{lightcurve}
\end{figure}
\vskip 0.25 in

Once the peak brightness has been determined, the supernovae are plotted on a Hubble diagram (brightness vs. redshift) as in Fig. \ref{hubblediag}.  Redshift is a measurement of the expansion since light left the SN and brightness is a measure of the time since light left the SN.  Shown in Figure \ref{hubblediag} are a sample of  fully analyzed distant ($ z > 0.3$) Type Ia supernovae from the HZT and SCP as well as a sample of  nearby ($ z < 0.1$) SNe Ia from the Calan-Tololo Survey.  Superposed are three cosmological models,  the standard model ($\Omega_{M} = 1$, no cosmological constant), an open model ($\Omega_ {M} =0$, $\Lambda =0$) and a $\Lambda$ model ($\Omega_{M}$, $\Lambda > 0$).  The SNe data favor a cosmology with low matter density and a cosmological constant, $\Lambda$, or dark energy model.

\begin{figure}
\vskip 0.25 in
\caption{Top panel: The Hubble diagram for type Ia supernovae
combining data from the Calan-Tololo Survey (red) and CfA Supernovae
(blue) at low $z$ and the SCP (red) and HZT (blue) at high redshift.
Three possible cosmologies are shown with lines. A flat, matter
dominated universe is the dashed line, an open, matter dominated
cosmology as a dotted line and a flat, cosmological constant
dominated universe is indicated by a solid line. Bottom panel:
same as the top panel but with the open, matter dominated universe
subtracted from the data. A lambda dominated universe is clearly
favored by both research group's results.}
\label{hubblediag}
\end{figure}
\vskip 0.25 in

These data can be jointly fit for mass density and vacuum energy density (cf. \cite{goobarperlmutter}).  Figure \ref{confidence} shows the confidence region (in blue)  of the joint fit for SCP supernovae, yielding a "best" value of $\Omega_{M} \sim ~0.3$, and $\Omega_{\Lambda} \sim~ 0.7$ (\cite{perlmutteretal1999}).  The HZT also obtained similar results (\cite{riessetal1998}). Drawn in this figure are the confidence regions from the CMB (MAXIMA, Boomerang) experiments ( \cite{langeetal}, \cite{balbietal}) - perpendicular to that of the SNe, and from cluster counting programs (for example, \cite{bahcalletal1999}).  
The supernovae evidence for a vacuum energy density  is in remarkable concordance 
with the galaxy cluster measurements (sensitive only to the value of  $\Omega_{M}$) (\cite{bahcalletal1999}), and the recent Cosmic Microwave Background results (e.g Maxima, 
Boomerang) which are sensitive to curvature, $\Omega_{\kappa}$ (\cite{langeetal}, \cite{balbietal}). 

\begin{figure}
\vskip 0.25 in
\caption{The two-parameter joint fit confidence regions for the vacuum energy density and mass density.  The oval contours (upper left) are the fit for a sample of high redshift type Ia supernovae.  The results from the recent cosmic microwave balloon experiments lie along the flat universe line, perpendicular to the SNe contours.  Also included are the confidence regions for various cluster counting methods, which are insensitive to $\Lambda$.}
\label{confidence}
\end{figure}
\vskip 0.25 in

It is this result that leads to a surprising conclusion.  The Universe's expansion is 
accelerating rather than merely decelerating as would be expected due to gravity alone.  
This means that the standard cosmology model - matter dominated, flat universe --  
appears to be wrong, leading to the conclusion that our understanding of fundamental 
physics is incomplete.  We are left with primary questions:  What is the nature of the 
vacuum energy density that accelerates the expansion?  When was the ``era of 
deceleration"? What is the history of expansion of the universe?   
 
One approach  to investigating the nature of this new energy component is to determine 
its equation of state.  A cosmological constant has an equation of state $w = p/{\rho} = -1$.  If 
this vacuum energy is due to some other primordial field,  e.g quintessence (see discussion below) or 
cosmic defects  its dynamical properties would be very different from a 
cosmological constant.  In Figure \ref{wm}, we show the best fit confidence regions from the 
supernova work  in the $\Omega_{M}$ -  $w$  plane for a vacuum energy density whose 
equation of state is $w = p/{\rho}$, constrained to a flat universe.   The value for $\Omega_{M}$ is between 0.2 and 0.4, and $w < -0.33$ (cf. \cite{perlmutteretal1998}, \cite{garnavichetal1998}).  Also shown are the regions covered by quintessence tracker modesl (e.g. \cite{albrecht}), cosmic 
strings  and a cosmological constant ($w=-1$).  The small (yellow) area within the 
current SN confidence regions is the expected confidence region allowed by the proposed SNAP satellite
observations for $w = -1$ and $\Omega_{M} = 0.28$.

\subsection {The Future:  Dark Energy Parameters and SNe Ia} 
 
A proposed experiment to achieve the precision required to determine the values of the 
cosmological parameters and  investigate the dark energy  properties  is  with SNAP's 
(Supernova Acceleration Probe) supernova program (\cite{snowmass2001}).  By exploiting and extending the 
brightness-redshift relation of SNe Ia to higher redshifts, we can probe the era of 
deceleration as well as measure the actual magnitude of the values for $\Omega_{M}$ and  
of $\Omega_{DE}$.  Increasing  the number of  discovered and studied SNe Ia to $\sim$ 2500  
per year, out to redshifts of $z = 1.7$, will  enable this technique to  identify and limit 
systematic errors while satisfying the statistical requirements. The recent results of SN 
1997ff, a SN Ia discovered in the Hubble Deep Field (\cite{riessetal2001}) demonstrates
 the power of 
this kind of experiment.   With such a dataset, $\Omega_{M}$ and $\Omega_{DE}$ can be 
determined to a few percent and models for the vacuum energy tested (Figure 
\ref{wm}).  Perhaps then the missing energy problem may be understood. 

\begin{figure}
\vskip 0.25 in
\caption{In this figure we show the confidence region for high redshift supernovae in the $w$--$\Omega{M}$ plane for a flat universe with constant $w$, where $w$ is the equation of state parameter.  Also shown are the regions 
predicted by cosmic strings (at $w=1/3$), by a  range of quintessence models ($-0.8 < w < -0.3$) and by 
a cosmological constant ($w=-1$).}
\label{wm} 
\end{figure}
\vskip 0.25 in
 
\subsection {The Missing Energy Problem} 
 
The total energy density of the 
Universe is much greater than the 
energy density contributed by all baryons, neutrinos, photons, and dark matter. 
Deepening this mystery are the recent observations of type Ia supernovae which 
suggest that the expansion rate of the Universe is accelerating. One possible resolution is a cosmological 
constant which fills this energy gap. 
 
The existence of a cosmological constant, $\Lambda$,  at an energy density $\sim 120$ orders 
of magnitude below reasonable estimates presents a distinct challenge 
to fundamental physics. Adding to the challenge is the apparent coincidence (in 
cosmological scales) in the present-day amplitude of the missing energy, dark 
matter, baryon, and radiation densities. If there is a significance to this 
multiple coincidence, then we may suspect that there is some mechanism 
responsible (not unlike supersymmetry, which is theorized to pull together the 
gauge couplings in a triple ``coincidence''). This provides some motivation to 
consider a dynamical component for the missing energy. Hence, a logical alternative to the cosmological constant is ``quintessence,'' 
a time-dependent, spatially inhomogeneous, negative pressure energy component 
which drives the cosmic expansion. (See 
\cite{Ratra:1988rm,Peebles:1988ek,Frieman:1995pm,Coble:1997te,Caldwell:1998ii}.) 
 
\section{Quintessence} 
 
Quintessence (Q) is a time-varying, spatially-inhomogeneous, negative pressure 
component of the cosmic fluid. It is distinct from $\Lambda$ in that it is {\it 
dynamic}: the Q energy density and pressure vary with time and is spatially 
inhomogeneous. A typical example of quintessence is a scalar field slowly 
rolling down a potential, similar to the inflaton in inflationary cosmology. 
Unlike $\Lambda$, the dynamical field can support long wavelength fluctuations 
which leave an imprint on the CMB and the large scale distribution of matter. 
Another, critical distinction is that $w$, the ratio of the pressure ($p$) to 
the energy density ($\rho$), is $-1 < w \le 0$ for quintessence, whereas $w$ is 
precisely $-1$ for $\Lambda$. Hence, the expansion history of the Universe for 
$\Lambda$  and Q models are different. There is much rich behavior to explore 
in a cosmological model with quintessence. (See \cite{Wang:2000fa} for an 
exhaustive survey of cosmological phenomena impacted by quintessence.) 
 
Fundamental physics, {\it e.g.} those theories of gravity and fundamental 
interactions beyond the standard model of particle physics, provide some 
motivation for light scalar fields, one of which may serve as a cosmic Q 
field.  In this way, quintessence serves as a bridge between the fundamental 
theory of nature and the observable structure of the 
Universe. 
 
In the following, we identify the cosmological parameters which describe the 
dark energy in the framework of the quintessence scenario. In discussing these 
parameters, we have a mind towards both observation and the interpretation in 
terms of the underlying physics. 
 
\subsection{Background Parameters: $\Omega$ and $w$} 
 
The existence, and in fact dominance, of dark energy in the Universe is 
characterized by $\Omega_Q$, the energy fraction. While a broad range of 
observational evidence indicates a low matter density fraction, $\Omega_m < 1$, 
the existence of dark energy does not obviously follow. That the missing energy 
is filled by something, rather than nothing (perhaps curvature), is deduced 
after the critical cosmic microwave background measurement of the geometry of the Universe, and hence, $\Omega_Q \approx 1 - \Omega_m$. The present best 
limits on the dominating dark energy density give 
$$ 
\Omega_Q \approx 0.7 
$$ 
with typical $1 \sigma$ uncertainties of $\pm 0.1$, which grow when 
certain assumptions are relaxed (e.g. assuming flatness, $h$, $\Lambda$).  
(See \cite{Bahcall:1999xn} and \cite{BOOMresults}.) Because the 
dark energy does not cluster significantly (at least on scales $\lesssim 
50$~Mpc), we cannot assess its 
energy density in the same ways that we tally up the mass density of stars or 
galaxies. Nevertheless, because of the widespread role of the dark energy density,  numerous observations can help pin down this number.  For example, type 1a supernovae magnitudes, galaxy counts, the 
CMB, and strong gravitational lens optics all have in common a dependence on 
the relationship between luminosity distance and redshift, 
$$ 
d_L(z) = {c \over H_0} (1 + z) \int_0^z \, dz' / 
\sqrt{ (1+z')^3 \, \Omega_m  + \Omega_Q[z'] }. 
$$ 
(Other distance have different weights in redshift--- the differences are not important here, which is actually a disadvantage 
in measuring $w$ \cite{Maor:2001jy}.)  Assuming a simultaneous 
measurement of the Hubble constant, we see that distance measurements are 
sensitive to the dark energy density over a range of redshift. We also see 
that characterizing the dark energy density by a single parameter, $\Omega_Q$ 
at $z=0$, only makes sense if the energy density is slowly varying 
over the interval observed. 
 
To construct a cosmological model, we must know the time evolution of the dark 
energy density. Theory has offered up a variety of ideas, yet rather than get 
bogged down in the details of one particular model (and which one to choose?), 
a reasonable first step is to encode the time evolution in the equation of state, 
$w$, defined as the ratio of the pressure to the energy density of the 
homogeneous dark energy. A given time history for $w$ can be used to determine 
the history of the energy density, 
$$ 
\rho_Q(a) = \Omega_Q \rho_{crit} 
\exp{\left(3\int_a^1 [1 + w(a')]\, d\ln a'\right)},  
$$ 
where $\Omega_Q$ is the energy fraction at the present, when $a=1$ is the 
scale  factor. For a cosmological constant, $w=-1$ for all times, and the 
energy density is a constant. Whereas for a scalar field evolving in a 
potential, the field equations of motion must be solved in order to  follow the 
pressure and energy density, a time history for $w$ can suffice, and in fact 
implies a potential and field amplitude, given by the parametric equations 
\begin{eqnarray} 
V(a) &=&   \frac{1}{2}[1 - w(a)] \rho_Q(a) \cr 
Q(a) &=&   \int d a' {\sqrt{1 + w(a')} 
\over a' H(a')} \sqrt{\rho_Q(a')}. \nonumber 
\end{eqnarray} 
Reconstruction of a continuous function such as  the potential from 
observations of the expansion history is an unrealistic goal, considering 
systematic uncertainties. But this equivalence between the equation of state 
and the potential, $w(a) \leftrightarrow V(Q[a])$, immensely simplifies the 
study of quintessence. It means a theorist can forgo details such as masses and 
exponents when comparing with observation, and observers need only interpret 
phenomena in terms of an average $w$ over a range of red shift. This average, 
$$ 
{\Large\bar w} = \int da \, \Omega_Q(a) w(a) \,\, /  \int da \, \Omega_Q(a) 
$$ 
is the second parameter which we require to characterize quintessence 
or dark energy. The best $2\sigma$ limit on the equation of state gives 
$$ 
{\Large\bar w} \lesssim -0.5 
$$ 
based on a conservative interpretation of the current data 
\cite{Wang:2000fa,Garnavich:1998th,Perlmutter:1999jt,Efstathiou:1999tm}. 
 
Introducing ${\Large\bar w}$ enlarges our cosmological parameter space, and 
reduces the sharpness of existing measurements of the other parameters. It is 
important to realize that theoretical models of quintessence are not evenly 
distributed in ``$w$-space.'' This should be taken as a warning to all those 
statisticians desiring to marginalize over $w$ (a theorist's prior is 
inhomogeneous at best!). Furthermore, the physical, observable difference 
between two cosmological models with a fixed difference in $w$, $\Delta w$, is 
not uniform throughout ``$w$-space.'' For example, using the CMB as a 
discriminant, the difference between $w=-1$ and $w=-0.9$ cosmologies is much 
smaller than the difference between $w=-0.6$ and $w=-0.5$. An alternate 
description of the effect of the dark energy on the cosmic expansion may be 
made with the red shift at which the quintessence and matter densities were 
equal, $z_{QM}$, or the red shift at which acceleration began, $z_{ACC}$. These 
red shifts can be used to describe the change of behaviour in, say, the growth 
rate of matter density perturbations or the accumulation of the integrated 
Sachs-Wolfe effect in the CMB.  Investigations of various methods to measure 
$w$  include \cite{Newman:1999cg,Huterer:2000mj,Haiman:2000bw}. Probably the 
most ambitious, focused method for attacking the dark energy, with the 
potential for dramatic results, is the Supernova Acceleration Probe, or SNAP 
(see {\tt snap.lbl.gov} and elsewhere in these proceedings). 
 
It is useful to make some comments on $w$ in the context of quintessence. 
Among the field theory models proposed, few predict a constant equation of 
state when the cosmic scalar field comes to drive the expansion rate. 
Technically, the change in the Hubble damping term pushes the field into a 
transient regime which, because the Universe is not completely Q-dominated, 
still persists today. However, for a slowly evolving field, $\dot w \lesssim 
H$, the average ${\Large\bar w}$ is not too different from the range of values 
sampled by $w(a)$ over the last Hubble time. On the other hand, models have 
been proposed which feature an oscillating field in which case averaging 
smooths out rapid oscillations ($\dot w \gg H$) and ${\Large\bar w}$ describes 
the slow drift in the pressure which drives the accelerating expansion. 
 
What will measurement of $w$ tell about the theory? In the case ${\Large\bar w} 
\approx -1$ it will be impossible to discriminate between whether the dark 
energy is a true cosmological constant or just a very slowly evolving field, 
unless there exist nonlinear excitations of the $Q$ field. If ${\Large\bar w} 
\lesssim -0.7$ we can conclude that the dark energy is not a quintessence 
tracker field \cite{Steinhardt:1999nw}. If a measurement of ${\Large\bar w}$ at 
an earlier red shift interval is made, then we could infer the sign of the 
derivative, $\dot {\Large\bar w}$. If $\dot{\Large\bar w}<0$ the equation of 
state is becoming more negative, driving greater acceleration; this is a common 
feature of slowly rolling fields, especially trackers \cite{Zlatev:1999tr}. If 
$\dot{\Large\bar w}>0$ the equation of state is becoming less negative, 
suggesting the period of acceleration is temporary, a coasting stage. Any of 
these results would have a tremendous impact on theory. 
 
The lower bound on $w$ is often taken to be $-1$, the limiting case of a 
cosmological constant. However, there is no strict prohibition on a more 
negative equation of state, $w < -1$. Certainly, a canonical, homogeneous 
scalar field in Einstein gravity cannot achieve such super-negative pressure, 
and violation of certain of the energy conditions ($\rho + p < 0$ with $\rho > 
0$ violates the dominant energy condition) may signal unusual physics. Indeed, 
toy models of quintessence have been constructed which have a super-negative 
equation of state \cite{Caldwell:1999ew}. However, the interpretation of 
observation and experiment should not be constrained by these considerations. 
After all, nature may be trying to tell us something.

\subsection{Perturbative Parameters: $\lambda$ and $c$} 
 
A time-evolving dark energy component must necessarily fluctuate in response to 
inhomogeneities in the surrounding matter distribution. Hence, we can expect 
there to be fluctuations in the quintessence which leave an imprint on the 
large scale distribution of matter and the cosmic microwave background. 
Detection of these fluctuations would reveal information about the microphysics 
of the dark energy. 
 
The microphysical model describes the behavior of density and pressure 
fluctuations, and any other properties of the dark energy. This is information 
not contained in the equation of state, which merely reflects the gross 
features of the medium. Nor is this information contained in the formal theory 
of cosmological perturbations, which provides evolution equations for certain 
geometric quantities. What we require is a description of excitations in the 
dark energy, giving the relationship between $\delta P$ and $\delta\rho$ as a 
function of time and length scale. In general, this requires a continuous 
function of two variables for the speed of propagation of  statistically 
homogeneous perturbations in the dark energy, $c_s^2(t,\vec x) = \delta P / 
\delta\rho$. As a simplification, we introduce two new cosmological parameters, 
$\lambda_Q$ and $c_Q$ which may be approximately understood as  $c_s = 1$ for 
$\lambda < \lambda_Q$ and $c_s = c_Q$ for $\lambda > \lambda_Q$. 
 
In the case of quintessence, the microphysical model is determined by the 
Lagrangian density 
$$ 
L = \frac{1}{2} \partial_\mu Q\, \partial^\mu Q - V(Q) 
$$ 
for the cosmic scalar field $Q$ with a potential $V(Q)$. Fluctuations $\delta 
Q$ about the homogeneous background field $Q$ obey the wave equation 
$$ 
\delta Q'' + 2 {a' \over a} \delta Q' + \left( V_{,QQ} - \nabla^2 \right)\delta Q 
= -{1\over 2} h' Q' 
$$ 
from which the solution is used to construct quantities such as the energy 
density, which is linear in $\delta Q,\, \delta Q'$. Here, $h$ is the trace of 
the synchronous gauge metric perturbation, and for practical purposes is 
proportional to the time derivative of the matter density contrast. So we can 
see that the quintessence reacts to the external gravitational field through 
$h$. The nature of the response is determined by $V_{,QQ}$, which characterizes 
the effective mass of the scalar field, $m_Q = \sqrt{V_{,QQ}}$. The speed of 
propagation of the perturbations is $c_Q = 1$, so that the characteristic 
perturbation length scale is $\lambda_Q = 1/\sqrt{V_{,QQ}}$. Taking advantage 
of the equivalence between the equation of state and potential for the scalar 
field, one can express $V_{,QQ}$ in terms of $a, \, w$ and time-derivatives 
thereof.  For a slowly evolving field, it turns out that $V_{,QQ} \propto H^2$, 
meaning that fluctuations $\delta Q$ on scales smaller than the Hubble scale 
dissipate, keeping the field smooth, non-clustering. On larger scales, the 
field is unstable to gravitational collapse and long wavelength perturbations 
develop. These perturbations leave an imprint on the large scale distribution 
of matter. 
 
There are variations on the basic quintessence model which describe the same 
background, characterized by parameters $\Omega_Q$ and $w$, but have very 
different microphysics and in turn, different clustering properties. One 
example is k-essence \cite{Armendariz-Picon:2000dh,Armendariz-Picon:2001ah}, 
a scalar field with a non-canonical kinetic term in the Lagrangian, for which 
the sound speed can be small, $c_Q \ll 1$, allowing for pressureless 
fluctuations to collapse via gravitational instability. Another example is a 
rapidly oscillating field, for which $\dot w$ can be much larger than $H$, in 
which case the clustering scale can be much smaller that the Hubble length, 
$\lambda_Q \ll H^{-1}$, while keeping ${\Large\bar w}$ sufficiently negative to 
drive the accelerated expansion \cite{Sahni:2000qe,Dodelson:2000fq}.  
 
The best shot at detecting the long wavelength fluctuations predicted in 
quintessence models is to use full sky maps that trace cosmic structure on the 
largest scales. Cross-correlation of different tracers, such as the CMB and 
weak lensing, can be used to isolate the unique features of quintessence.  (See 
the discussion in \cite{Caldwell:2000wt}.) This may be difficult, however, 
since large scale phenomena is fundamentally limited by cosmic variance, not to 
mention systematic uncertainties.  Yet the payoff would be worth it. Detection 
of dark  energy clustering at lengths below the Hubble scale would provide an 
important clue to the physics of dark energy, indicating a solution different 
from the cosmological constant. 
 
\subsection{Circumstantial Evidence: $\dot G/G$ and $\dot\alpha/\alpha$} 
 
Detection of a slow drift in the value of the gravitational Newton's constant, 
or the electromagnetic fine structure constant, would indicate radical new 
physics well outside the confines of conventional theory. Such a result might 
also be a signal for an evolving cosmic scalar field:  scalar fields which 
modulate the strength of coupling constants appear frequently in the same 
models of high energy physics which motivate models of quintessence. With a 
coupling to the gravitational or electromagnetic action, such scalar-tensor or 
scalar-vector theories with a flat potential could also provide the dark 
energy. (See \cite{Amendola:2000er,Baccigalupi:2000je} and references therein.) 
Because constraints on the evolution of such constants are so tight (see 
\cite{Carroll:1998zi} for a review in the quintessential context), the 
quintessence coupling must be relatively insensitive to the field  evolution, 
or the field $Q$ must be very slowly creeping down the potential. Just the sort 
of behavior needed for negative pressure.

\subsection{Dark Energy Summary} 
 
Provided that the trend in the observational data continues, we will find 
ourselves with strengthening evidence for the missing energy problem. How do we 
then proceed? 
 
The first order of business is to refine the measurements of the basic 
cosmological parameters. That is, we must verify that the matter density is 
low, $\Omega_m < 1$, that the spatial curvature is negligible, $|\Omega_k| \ll 
1$, and that there is indeed a missing energy problem. Since most phenomena are 
sensitive to a combination of cosmological parameters, the measurement of the 
Hubble constant must also be further refined. 
 
Given that the missing energy problem is real, the next logical step will be to 
characterize the equation of state, $w$, in order to determine whether the dark 
energy is $\Lambda$, $Q$, or other.  For fundamental physics,  $\Lambda$ and 
$Q$ represent new, ultra-low energy phenomena beyond the standard model. If 
firmly established by observations and experiment, the discovery will go down 
in history as one of the greatest clues to an ultimate theory. 
 
Once the basic properties of the dark energy are determined, $\Omega_Q$ and 
$w$, we can begin to ask questions about the microphysics --- what is it? Long 
wavelength fluctuations, manifest in very large scale structure and the CMB are 
potential clues. 
 
Finally, the structure of our cosmological models, the missing energy problem 
and the quintessence hypothesis, are predicated on the validity of Einstein's 
GR, and the existence of cold dark matter supporting a spectrum of primordial, 
adiabatic density fluctuations. It is critical that we test this framework at 
the same time that we look for the dark energy. 
 
\section{Large Scale Structure: Theory} 
 
Large scale structure can constrain the cosmological parameters, through the effect of density fluctuations.  It is impossible to do justice to the vast field of large 
scale structure studies in a few pages in this report. 
Instead, we choose to focus on two issues: how secure is the gravitational instability 
paradigm for structure formation? and  what does large 
scale structure teach us about particle physics? emphasizing a review of the conceptual issues 
rather than detailed quantitative estimates and measurement.   We will also discuss several 
new techniques/observational windows in the context of the 
first question. (Traditionally, the term large scale structure 
is reserved for the study of the large scale distribution of mass 
as manifested through galaxies. Here, we use it to include other tracers 
as well).  We refer the reader to 
excellent reviews in \cite{peebles80,roman01} for more discussions.

\subsection{The Gravitational Instability Paradigm} 
 
The most successful and the most thoroughly studied theory of structure formation 
postulates that the observed large scale structure, as manifested through 
galaxies, clusters, weak gravitational lensing as well as quasar absorption, 
arise from the growth of fluctuations under gravity \cite{textbooks}. 
A common and necessary ingredient of such a theory is the presence 
of a large amount of dark matter which interacts very weakly except 
through gravity. How successful is this theory? 
 
Tests of this theory can be roughly divided into three categories. 
As we will see, none of them offer a completely clean test, but 
together they form a compelling, though far from air-tight, 
argument in favor of gravitational instability. 
 
\subsubsection{The Clustering Hierarchy} 
 
Assuming Gaussian initial conditions (such as predicted or at least 
preferred by inflation), 
gravity predicts a definite hierarchy of correlations where 
$\xi_N \propto {\xi_2}^{N-1}$ where $\xi_N$ is the N-point correlation 
function of the mass \cite{peebles80,fry,davis}. The proportionality constant 
can be robustly predicted on large scales using perturbation theory 
\cite{bernardeau}. 
Such a hierarchy (including the predicted 
proportionality constant) appears to be consistent with the observed 
galaxy correlations \cite{enrique,istvan,enriquelam,romantrispec}. However, 
one uncertainty remains in the interpretation: the relation between 
mass and galaxy, the bias. To take a simple example, 
suppose the fluctuation in galaxy density ($\delta_g = (n_g - \bar n_g) / \bar n_g$, where 
$n_g$ is the galaxy density and $\bar n_g$ its mean) 
is related to the fluctuation in mass density ($\delta = (\rho - \bar \rho)/ \bar \rho$) 
via: 
\begin{equation} 
\delta_g = b \delta 
\label{bias} 
\end{equation} 
where $b$ is the bias factor (see \cite{fryenrique,mowhite,pendekel} for 
more general biasing schemes). Then, the observed galaxy skewness, which 
relates the spherically averaged three-point function to the 
averaged two-point function i.e. $S^g_3 \equiv \langle \delta_g^3 \rangle / 
\langle \delta_g^2 \rangle^2$ 
is given by 
\begin{equation} 
S^g_3 = S_3 / b 
\label{Sgb} 
\end{equation} 
where $S_3$ is the mass skewness, defined by $S_3 
\equiv \langle \delta^3 \rangle / \langle \delta^2 \rangle^2$. 
Inference of the mass hierarchical constants (such as $S_3$), which 
is predicted by theory, from the observed galaxy distribution 
therefore depends on assumption about how galaxy traces mass. 
What is perhaps surprising is that the theoretically predicted 
$S_3$ (analogous higher order quantities) does appear to agree with 
the inferred $S_3$ if bias is assumed to be,  $b \sim 1$ \cite{enrique,istvan}. 
 
That gravitational instability, together with the simplest but by no means 
obvious assumption that galaxies trace mass, can readily explain the 
observed hierarchy should be counted as one successes of the theory. 
However,the interpretation is somewhat complicated by the issue of bias. 
Workers in the field have turned the argument around: assuming gravitational 
instability, use the observed hierarchy to constrain bias \cite{frytrispec,romantrispec}. 
See also \cite{steidel} for evidence that bias is non-trivial at high redshifts 
($z \sim 3$), and \cite{frybias,maxpeebles} for discussions of how bias might evolve under 
simple assumptions. 
 
\subsubsection{Velocities} 
 
Another robust prediction of gravitational instability is the relation 
between density and velocity. Conservation of mass tells us \cite{bertschinger}: 
\begin{equation} 
{\partial \delta / \partial \tau} = -\nabla \cdot v 
\end{equation} 
where we have kept terms to first order (i.e. $\delta$ and peculiar velocity 
$v$ are both small), and $\tau$ is the conformal time. 
According to linear perturbation theory, the fluctuation 
$\delta$ grows as $\delta \propto D(\tau)$ where $D(\tau)$ is 
the growth factor dictated by gravity. Hence the above equation can be 
rewritten as 
\begin{equation} 
\delta = - ({D/D'}) \nabla \cdot v 
\end{equation} 
where $'$ denotes derivative with respect to $\tau$. 
Such a simple relation between overdensity and peculiar velocity can 
be tested, using galaxies as test particles. 
There are many ways to do so, but they fall roughly into 
two classes. The first makes direct use of peculiar velocities measured 
in surveys to reconstruct the density field, which is then compared with the observed density field 
or vice versa \cite{potent,dekelpotent}. This approach requires accurate distance indicators, however. The commonly used Tully-Fisher method gives error-bars that are somewhat large; 
independent, more reliable methods would be desirable \cite{riessvel,SBF}  The second  makes use of the fact that 
peculiar velocities introduce a systematic distortion of the galaxy 
density field and its correlations in the observed redshift-space \cite{kaiser,fisher,hamilton}. 
 resulting in  a 
systematic line-of-sight-squashing of the correlation function on large scales \cite{kaiser}. 
 
Much research has been carried out following the above two approaches. 
The quantitative success of these tests depends on two additional 
parameters: 1. $\Omega_m$ (the matter density in units 
of the critical density) which largely determines the growth rate 
$D$, and 2. the bias $b$, because, as before, galaxy density is observed 
instead of the mass density. Indeed, very often, the problem is turned around: 
instead of testing gravitational instability, one assumes it and uses 
the above to measure $\Omega_m$ and $b$ or some combination of them. 
The fact that such measurements yield values for $\Omega_m \sim 0.3$ and 
$b \sim 1$ \cite{velocityrecent,zrecent}, 
which are consistent with other independent measurements (see 
other sections in this report), should be counted as another success of the 
theory. However, it remains to be seen whether improving accuracy will 
reveal inconsistency. 
 
\subsubsection{Fluctuation Growth} 
 
The rate at which fluctuations grow via gravitational instability, 
from the epoch of recombination (probed by the microwave background) to 
the present, can be predicted precisely provided we know the amount of dark matter, 
the nature of dark matter (whether it is hot or cold), the primordial power 
spectrum (particularly its scale dependence), and so on. Gravitational instability plays the role of bridging the gap between 
information from early epochs to late epochs. As such, 
microwave background and large scale structure data test gravitational instability 
only to the extent that some of the parameters can 
be independently constrained, or have values that are strongly motivated by 
theory (i.e. inflation). 
Indeed, the combination of microwave background and large scale structure, 
the latter as measured from surveys of present day galaxies, has provided 
a wealth of constraints on several cosmological parameters \cite{boompapers}. 
 
Methods of measuring fluctuation growth include measuring the galaxy correlation as a function of redshift (though growth can be confused with bias evolution) and counting of the number density of clusters as a function of redshift. Clusters as rare events probe the tail of 
the fluctuation probability, and provide a sensitive measure of 
the amplitude of the power spectrum and its evolution (the latter 
determined by $D(\tau)$ and hence by $\Omega_m$) \cite{clusters}. 
It is also a useful diagnostic of non-Gaussianity in the initial fluctuations 
\cite{clustergauss}.   The  Lyman-alpha forest 
or weak gravitational lensing can be used to more directly measure 
the evolution of mass clustering. 
 What is impressive about the results using these methods 
is that the simplest assumptions about the various 
components seem to hold up well e.g. cold dark matter with a scale invariant 
power spectrum of adiabatic, Gaussian fluctuations in a flat universe -- together with 
growth via gravitational instability. The major surprise in this picture 
is the need for a cosmological constant or some form of dark energy, as 
independently confirmed by supernovea measurements \cite{SN}. 
Expected improvements in precision will severely test this paradigm \cite{baryon}.

\subsubsection{A Note on Galaxy Formation} 
 
The growth of fluctuations with a power spectrum $P(k) \sim k^n$, where $n > -3$ 
and $k$ is the wavenumber, 
leads to a generic prediction: small objects go nonlinear first as one 
marches forward in time. This is easy to see: the root-mean-squared 
amplitude of the mass fluctuation at a given wavelength $\lambda = 2\pi/k$ behaves as: 
\begin{equation} 
\sqrt{\langle |\delta_k |^2 \rangle}\sim \sqrt{k^3 P(k)} \propto D(t) k^{(3+n)/2} 
\end{equation} 
where $\delta_k$ is a Fourier mode, and $D(t)$ is the fluctuation growth rate. 
Clearly, if $n > -3$, wavemodes with a high $k$ or small wavelength go 
nonlinear first. This leads to a picture of heirarchical galaxy formation in which small objects collapse first and then subsequently merge to form larger nonlinear objects. 
As dark matter collapses, gas accretes and is shocked, eventually cooling to 
 to form stars and galaxies \cite{whiterees}. 
The theory of hierarchical galaxy formation has developed into a major industry 
\cite{whitereview}. 
Because of the inherently complex gas physics, robust and precise predictions are 
hard to come by, except on large scales where it is believed that 
a simple linear bias (eq. \ref{bias}) suffices to describe the galaxy two-point 
correlations (a generalization of the linear bias into a power series 
is needed to describe the higher order correlations; see \cite{fryenrique}). 
The most tractable part of the problem: namely, 
the formation and merging of dark matter halos, has been very successfully 
solved by both numerical N-body simulations \cite{edNbody} 
and analytic excursion set theory \cite{excursion}. 
Recently, there has been some excitement over the development of 
a formalism that parametrizes in an economic way our ignorance of 
the detailed gas physics, and allows us to make progress towards 
linking small-scale clustering to galaxy formation physics \cite{halomodel}.  
The search for robust predictions from the hierarchical galaxy formation theory 
remains an important goal. For possible problems with this theory, see \cite{peebles}.  
 
\subsection{The Particle Physics Connection} 
 
One of the main goals of large scale structure studies is to  
discover the nature of the density fluctuations in the early universe. 
This is where large scale structure and particle physics meet. 
At present, the most comprehensive and successful theory of 
the initial conditions of the universe is inflation \cite{inflation}. 
The simplest models of inflation predict a spectrum of  
nearly scale invariant, adiabatic and Gaussian density fluctuations -- 
which is remarkably consistent with observations so far \cite{hu}.  
The nature of these fluctuations are tied to the form of the inflaton 
potential \cite{potential}, hence precision measurements of the fluctuations through 
the microwave background and large scale structure will provide important 
clues to physics at the energy scale of inflation (the canonical scale 
is the GUT scale). Several remarks are in order. 
 
\begin{itemize} 
 
\item One of the key parameters to measure is the power spectral index and 
perhaps its possible running with scale. One should have as long a lever arm as 
possible for such a measurement -- in other words, different measurements 
probe the power spectrum at different scales, and it is important to use 
all measurements, from the microwave background and galaxy surveys to 
weak-lensing and the Lyman-alpha forest, to cover as wide a range in scales 
as possible.  
 
\item Constraints on possible  
non-Gaussianity and non-adiabaticity of the fluctuations are at present  
quite weak. Many of the current constraints on cosmological parameters 
assume Gaussianity and adiabaticity of the initial conditions. Improved high order correlation 
measurements from the microwave background and large scale structure 
will be very helpful in constraining non-Gaussianity \cite{romanverde}. 
Adiabaticity is harder: while pure isocurvature fluctuations are 
not viable \cite{texture}, models where adiabatic and isocurvature 
fluctuations are mixed together are difficult to rule out. Microwave 
background polarization  
measurements will help break some of the degeneracies  
\cite{bucher}. 
 
\item One should keep in mind the possibility new physics that takes over 
at high energies or at short distances might leave an imprint in the 
quantum fluctuations which are stretched by inflation to become the large 
scale structure. Suggestions include dynamical effects due to large extra 
dimensions \cite{extradim}, and novel vacuum effects \cite{transplanck}.  
 
\item The processing of primordial fluctuations depends crucially on  
the nature of the dark matter. That is why large scale structure provides 
interesting constraints on neutrinos (particularly their mass) because 
they behave as hot dark matter, suppressing fluctuations 
on small scales \cite{tegmark}. 
Interestingly, there has been some recent suggestions 
cold dark matter might fail to explain the observed nonlinear structures 
on small scales, such as the abundance of small mass halos, and the 
inner density profiles of galactic halos \cite{spergstein}. It is yet 
unclear if these problems might be resolved by astrophysical means \cite{navarro}.  
 
\end{itemize}

\section{Large Scale Structure: New Observational Windows} 
There are two recurring themes in the above discussion. One is that 
the inference of the mass correlations from the observed galaxy 
distribution is subject to the uncertainty of bias.  The other is that 
having multiple independent checks is extremely important in 
establishing the gravitational instability in general, and the 
inflation + Cold Dark Matter (CDM) framework in particular. This 
naturally brings us to the topic of new techniques to probe the mass 
fluctuation. In this section, we review two new techniques, weak 
lensing and the Lyman-alpha forest, which have recently emerged as 
powerful probes of large-scale structure. 
 
\subsection{Weak Lensing by Large-Scale Structure} 
 
Gravitational Lensing provides a unique method to directly map the 
distribution of dark matter in the universe. It relies on the 
measurement of the distortions that lensing induces in the images of 
background galaxies (for reviews see \cite{bar99, mel01}). This method 
is now used routinely to map the mass of clusters of 
galaxies. Recently, weak lensing has been statistically detected for 
the first time in random patches of the sky 
\cite{bac00,wit00,van00,van01,kai+00,rho99, mao01,hae01}. Unlike other 
methods which probe the distribution of light, this 'cosmic shear' 
technique directly measures the mass and thus the distribution of dark 
matter in the universe. Weak lensing measurements can can thus be 
directly compared to theoretical models of structure formation, thus 
opening wide prospects for cosmology. 
 
\begin{figure} 
\scalebox{0.5}
{\includegraphics{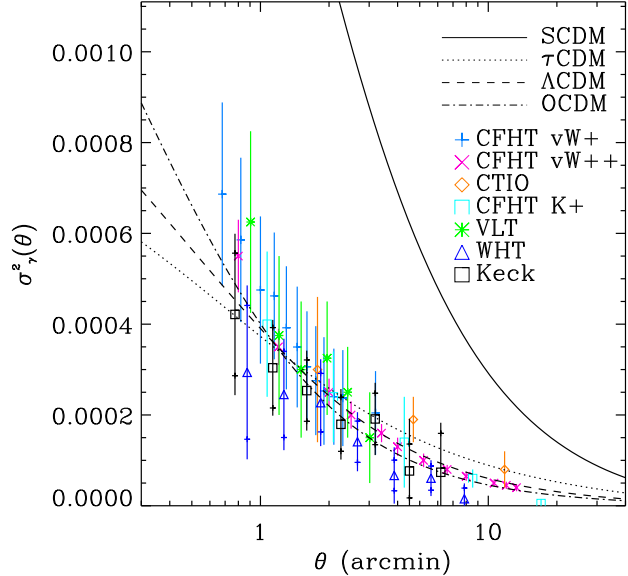}}%
\caption{The shear variance $\sigma_{\gamma}^{2}$ as a function of the
radius of a circular aperture measured by several groups.  The lines
represent the predictions of the different CDM models. The ground
based observations correspond to that of Wittman et al. (\cite{wit00},
CTIO), van Waerbeke et al. (\cite{van00}, CFHT vW+), Bacon et al.
(\cite{bac01}, WHT and Keck), Kaiser, Wilson \& Luppino
(\cite{kai+00}, CFHT K+), Maoli et al. (\cite{mao01}, VLT), and van
Waerbeke et al. (\cite{van01}, CFHT vW++). When applicable, the inner
error bars correspond to noise only, while the outer error bars
correspond to the total error (noise + cosmic variance). The
measurements are all consistent with the cluster normalised CDM
models.}
\label{variance} 
\end{figure}

As they travel from a background galaxy to the observer, photons get
deflected by mass fluctuations along the line of sight.  As a result,
the apparent images of background galaxies are subject to a distortion
which is characterised by the shear tensor $\gamma_{i}$. The shear
component $\gamma_{1}$ ($\gamma_{2}$) describes stretches and
compressions along (at $45^{\circ}$ from) the x-axis.  The simplest
statistics which can be used to characterise the resulting shear
patterns on the sky is the the shear variance
$\sigma_{\gamma}^{2}(\theta)$ in circular cells of radius
$\theta$. The measurements of $\sigma_{\gamma}^{2}(\theta)$ by the
different ground-based groups are shown in
Figure~\ref{variance}. Recently, cosmic shear has also been detected
from space using HST \cite{rho99,hae01} (not shown on the figure). The
agreement between the different groups is impressive given that they
used different telescopes and independent data analysis methods. The
curves in the figure show the predictions for several CDM models. The
measurements are in good agreement with cluster normalised models
($\Lambda$ CDM, OCDM and $\tau$CDM with $\sigma_{8}
(\Omega_{m}/0.3)^{0.5} \simeq 1$).
 
The amplitude and scale dependence of the cosmic shear signal depends 
strongly on the cosmological model. Measurements of the weak lensing 
power spectrum can thus be used to constrain cosmological parameters, 
such as $\Omega_{m}$, $\Omega_{\Lambda}$, $\sigma_{8}$, $\Gamma$, etc 
\cite{jai97,ber97,hu+99,van+98}. A precision of order of 10\% on 
these parameters can be achieved with weak lensing surveys of about 10 
square degrees. Such cosmic shear surveys can also be combined with 
CMB anisotropy measurements to break degeneracies present when the CMB 
alone is considered. This would yield improvements in the precision on 
cosmological parameters by about one order of magnitude \cite{hu+99}. 
 
\begin{figure} 
\scalebox{0.5}
{\includegraphics{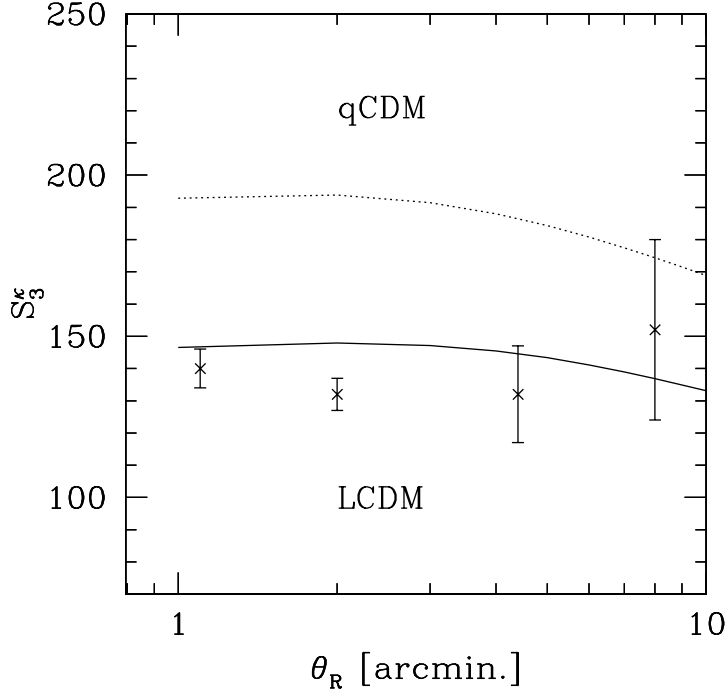}}
\caption{The solid and dotted lines show respectively the lensing 
convergence skewness as a function of angular scale for a $\Lambda$CDM 
(dominated by cosmological constant) model and for a qCDM model 
(dominated by quintessence or dark energy with equation of state: 
pressure/density $= -1/2$). Points with error-bars show simulated 
measurements with the expected statistical error for a $3^0 \times 
3^0$ weak-lensing survey with about $30$ galaxies per square 
arc-minute, where the galaxy median redshift is unity.  Note that 
$S_3^\kappa$ is largely independent of the details of the mass power 
spectrum. See \cite{huilens} for details. } 
\label{skewness} 
\end{figure}

The shear field is not gaussian, but, instead, has coherent structures 
on various scales. Bernardeau, van Waerbeke, \& Mellier \cite{ber97} 
have shown that the skewness $S_{3}^{\kappa}=\langle \kappa^{3} 
\rangle / \langle \kappa^{2} \rangle^{2}$ of the convergence field 
$\kappa$ (which can be derived from the shear field) can be used to 
break the degeneracy between $\sigma_{8}$ and $\Omega_{m}$, which is 
present when the shear variance alone is 
considered. Figure~\ref{skewness} shows the skweness as a function of 
angular scales expected in several cosmological models. 
 
The existing measurements described above are primarily limited by 
statistics. They will therefore be improved upon by future 
ground-based instruments, such as Megacam on CFHT \cite{bou00}, VISTA 
\cite{tay01}, LSST \cite{tys01}, or the novel WFHRI concept 
\cite{kai+00b}. From space, the upcoming camera ACS on HST, and, much 
more ambitiously, the future satellites SNAP \cite{per01} and GEST 
\cite{ben01} also offer exciting prospects. Broadly speaking, ground 
based measurements will cover large areas, while space-based surveys 
will yield higher resolution maps and reduced systematics thanks to 
the absence of atmospheric seeing. 
 
These future surveys will provide very accurate measurements of 
cosmological parameters through the measurement of the weak lensing 
power spectrum and higher-order statistics. They will also allow us to 
test some of the foundations of the standard cosmological model. For 
instance, the measurement of the power spectrum at different redshift 
slices and of the hierarchy of high-order correlation functions, will 
yield a direct test of the gravitational instability paradigm. Cosmic 
shear can also be used to measure the equation of state of the dark 
energy, and thus complement supernovae measurements in the 
constraining of quintessence models \cite{hut01, ben01, 
hu01}. Figure~\ref{skewness} shows how the skewness of the convergence 
field can be used to distinguish a $\Lambda$CDM model from a qCDM 
model. Cosmic measurements can also be used to test general relativity 
itself \cite{uza01}. 
 
For these instruments to yield to full promises of cosmic shear, a 
number of challenges have to be met. First, observationally, 
systematic effects, such as the PSF anisotropy, CCD non-linearities, 
etc, must be controlled and corrected for. From the theoretical point 
of views, calculations of the non-linear power spectrum, of high order 
statistics and of the associated errors must be improved to meet the 
precision of future measurements. The observational and theoretical 
efforts required to overcome these difficulties are very much worth 
while given the remarkable promisses which cosmic shear offers to 
cosmology. 
 
\subsection{Lyman-alpha Forest} 
 
\begin{figure} 
\scalebox{0.5}
{\includegraphics{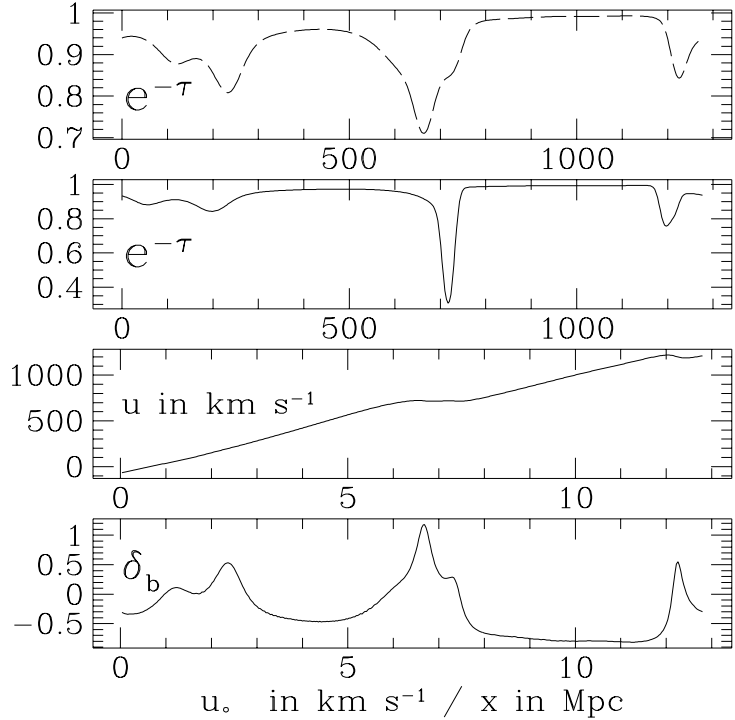}}
\caption{The bottom panel shows the baryon density fluctuation 
as a function of distance along a line of sight; the next 
panel up shows the total velocity (peculiar velocity plus  
Hubble flow) along the same. The top panel shows the 
corresponding $e^{-\tau}$, where $\tau$ is the 
optical depth and $e^{-\tau}$ is the probability of transmission,  
if peculiar velocities are ignored; the panel below shows the 
same when distortions due to peculiar velocities are taken into account. 
Note how $e^{-\tau}$, which is proportional to the observed flux 
of some quasar, provides almost a mirror image of the density profile, 
modulo peculiar velocities.} 
\label{forest} 
\end{figure}

Another emerging technique is the use of quasar absorption spectra 
at redshifts $z \sim 2 - 5$. The main idea is that  
fluctuations in (Lyman-alpha) absorption (known as the Lyman-alpha forest)  
can be directly related to fluctuations 
in the mass density, as illustrated in Fig. \ref{forest}.  
 
Numerical simulations and semi-analytical calculations show that 
the observed absorption fluctuations and several of their 
key characteristics can be attributed to mildly non-linear 
mass fluctuations ($\delta \sim 1$), which arise as a natural 
consequence of gravitational instability \cite{forestpapers}.  
In addition to allowing a measurement of the large scale structure 
and its growth at redshifts of a few \cite{croft}, quasar spectra also offer 
a geometrical method to measure the cosmological constant via 
the well-known Alcock-Paczynski effect \cite{ap}.  
The idea here is also quite simple. Consider a spherical ball 
placed at redshifts of a few. Its observed shape will be 
characterized by its redshift-extent and angular-extent. 
The presence of a cosmological constant turns out to distort 
the shape in a distinctive way, that cannot be mimicked by 
for instance spatial curvature: if the cosmological constant 
(or dark energy component with suitable equation of state 
to cause acceleration) is significant, the ball will appear 
squashed in the redshift direction.  
The abstract ``ball'' that the forest offers its the two-point correlation function 
of the absorption fluctuations, measured along sightlines 
to individual quasars, as well as across sightlines of close quasars 
\cite{apforest}.  
Such a test can be carried out 
with correlation functions of other objects as well, such as 
galaxies and quasars \cite{galQap}. At least one advantage 
of using the Lyman-alpha forest is that the mapping from 
mass fluctuation to absorption fluctuation is completely fixed 
by observations, allowing an estimate of distortions due 
to peculiar velocities, which must be accounted for in 
any application of the Alcock-Paczynski test using correlation 
functions \cite{velap}.  
 
\subsection{Conclusion} 
In summary, with the combination of new observational windows such as 
weak gravitational lensing and the Lyman-alpha forest, and the advent 
of large galaxy surveys, including the Two-degree-Field Survey and the 
Sloan Digital Sky Survey, the prospects of using large scale structure 
to constrain particle physics look brighter than ever. 
 
%
%
 
%
%
 
 

\end{document}